\documentclass{article}
\usepackage[left=1.25in,top=1.25in,right=1.25in,bottom=1.25in,head=1.25in]{geometry}
\usepackage{amsfonts,amsmath,amssymb,amsthm}
\usepackage{verbatim,float,url,enumerate}
\usepackage{graphicx,subfigure,epsfig,psfrag}
\usepackage{dsfont,bm,color,appendix}
\usepackage{natbib}
\usepackage{pdfpages}
\usepackage{algorithm}
\usepackage[export]{adjustbox}
\graphicspath{{figures/}}
\usepackage{hyperref}
\PassOptionsToPackage{hyphens}{url}\usepackage{hyperref}

\def\bzero{{\bf 0}}
\def\bone{{\bf 1}}

\def\bx{{\bf x}}
\def\by{{\bf y}}
\def\bz{{\bf z}}
\def\bX{{\bf X}}
\def\bZ{{\bf Z}}
\def\eps{\varepsilon}
\def\bbE{\mathbb{E}}
\def\calN{\mathcal{N}}
\newcommand\bbR{\mathbb{R}}

\providecommand{\minimize}{\mathop{\rm minimize}}

\newcommand{\norm}[1]{\left\|{#1}\right\|} 
\newcommand{\lone}[1]{\norm{#1}_1} 
\newcommand{\ltwo}[1]{\norm{#1}_2} 

\newtheorem{theorem}{Theorem}

\title{Feature-weighted elastic net: using ``features of features" for better prediction}
\author{ J. Kenneth Tay$^1$, Nima Aghaeepour$^{2,3,4}$, Trevor Hastie$^{1,4}$ and Robert Tibshirani$^{1,4}$}
\date{$^1$Department of Statistics, Stanford University\\
$^2$Department of Anesthesiology, Pain, and Perioperative Medicine, Stanford University\\
$^3$Department of Pediatrics, Stanford University\\
$^4$Department of Biomedical Data Sciences, Stanford University\\[2ex]
\today}

\begin{document}
\maketitle

\begin{abstract}
In some supervised learning settings, the practitioner might have additional information on the features used for prediction. We propose a new method which leverages this additional information for better prediction. The method, which we call the {\em feature-weighted elastic net}  (``fwelnet''), uses these ``features of features" to adapt the relative penalties on the feature coefficients in the elastic net penalty. In our simulations, fwelnet outperforms the lasso in terms of test mean squared error and usually gives an improvement in true positive rate or false positive rate for feature selection. We also apply this method to early prediction of preeclampsia, where fwelnet outperforms the lasso in terms of 10-fold cross-validated area under the curve (0.86 vs. 0.80). We also provide a connection between fwelnet and the group lasso and suggest how fwelnet might be used for multi-task learning.
\end{abstract}

\section{Introduction}

We consider the usual linear regression model: given $n$ realizations of $p$ predictors $\bX=\{x_{ij}\}$ for $i = 1, 2, \ldots, n$ and $j = 1, 2, \ldots, p$, the response $\by = (y_1, \ldots, y_n)$ is modeled as
\begin{equation}
y_i=\beta_0 + \sum_{j=1}^p x_{ij} \beta_j +\epsilon_i,
\end{equation}
with $\epsilon$ having mean $0$ and variance $\sigma^2$. The ordinary least squares (OLS) estimates of $\beta_j$ are obtained by minimizing the residual sum of squares (RSS). There has been much work on regularized estimators that offer an advantage over the OLS estimates, both in terms of accuracy of prediction on future data and interpretation of the fitted model. One popular regularized estimator is the elastic net \citep{Zou2005a} which minimizes the sum of the RSS and a combination of the $\ell_1$ and $\ell_2$-squared penalties. More precisely, letting $\beta=(\beta_1, \ldots,\beta_p)^T$, the elastic net minimizes the objective function

\begin{align}
J(\beta_0,\beta) &= \frac{1}{2}\ltwo{\by-\beta_0\bone -\bX \beta}^2+\lambda \left[\alpha \lone{\beta}+ \frac{1 - \alpha}{2}\ltwo{\beta}^2 \right] \label{eqn:elnet} \\ 
&= \frac{1}{2}\ltwo{\by-\beta_0\bone -\bX \beta}^2 + \lambda \sum_{j=1}^p \left( \alpha |\beta_j| + \frac{1 - \alpha}{2}\beta_j^2 \right). \label{eqn:elnet2}
\end{align}

The elastic net has two tuning parameters: $\lambda \geq 0$ which controls the overall sparsity of the solution, and $\alpha \in [0,1]$ which determines the relative weight of the $\ell_1$ and $\ell_2$-squared penalties. $\alpha = 0$ corresponds to ridge regression \citep{Hoerl1970}, while $\alpha = 1$ corresponds to the lasso \citep{Tibshirani1996}. These two tuning parameters are often chosen via cross-validation (CV). One reason for the elastic net's popularity is its computational efficiency: $J$ is convex in its parameters, which means that solutions can be found efficiently even for very large $n$ and $p$. In addition, the solution for a whole path of $\lambda$ values can be computed quickly using warm starts \citep{Friedman2010}.

In some supervised learning settings, we often have some information about the features themselves. For example, in genomics, we know that each gene belongs to one or more genetic pathways, and we may expect genes in the same pathway to have correlated effects on the response of interest. Another example is in image data, where each pixel has a specific position (row and column) in the image. We would expect methods which leverage such information on the features to perform better prediction and inference than methods which ignore it. However, many popular supervised learning methods, including the elastic net, do not use such information about the features in the model-fitting process.

In this paper, we develop a framework for organizing such feature information as well as a variant of the elastic net which uses this information in model-fitting. We assume the information we have for each feature is quantitative. This allows us to think of each source as a ``feature" of the features. For example, in the genomics setting, the $k$th source of information could be the indicator variable for whether the $j$th feature belongs to the $k$th genetic pathway.

We organize these ``features of features" into an auxiliary matrix $\bZ \in \bbR^{p \times K}$, where $p$ is the number of features and $K$ is the number of sources of feature information. Each column of $\bZ$ represents the values for each feature information source, while each row of $\bZ$ represents the values that a particular feature takes on for the $K$ different sources. We let $\bz_j \in \bbR^K$ denote the $j$th row of $\bZ$ as a column vector.

To make use of the information in $\bZ$, we propose assigning each feature a \textit{score} $\bz_j^T \theta$, which is simply a linear combination of its ``features of features". We then use these scores to influence the weight given to each feature in the model-fitting procedure. Concretely, we give each feature a different penalty weight in the elastic net objective function based on its score:

\begin{equation}\label{eqn:fwelnet-general}
J_{\lambda, \alpha, \theta}(\beta_0,\beta) = \frac{1}{2}\|\by-\beta_0\bone -\bX \beta\|_2^2+ \lambda \sum_{j=1}^p w_j(\theta) \left[\alpha |\beta_j | + \frac{1 - \alpha}{2} \beta_j^2 \right],
\end{equation}

where $w_j(\theta) = f \left(\bz_j^T \theta \right)$ for some function $f$. $\theta$ is a hyperparameter in $\bbR^K$ which the algorithm needs to select. In the final model, $\bz_j^T \theta$ can be thought of as an indication of how influential feature $j$ is on the response, while $\theta_k$ represents how important the $k$th source of feature information is in identifying which features are important for the prediction problem.

The rest of this paper is organized as follows. In Section \ref{sec:litreview}, we survey past work on incorporating ``features of features" in supervised learning. In Section \ref{sec:fwelnet}, we propose a method, the {\em feature-weighted elastic net} (``fwelnet"), which uses the scores in model-fitting. We then illustrate its performance on simulated data in Section \ref{sec:sim} and on a real data example in Section \ref{sec:realdata}. In Section \ref{sec:glasso}, we present a connection between fwelnet and the group lasso, and in Section \ref{sec:multitask}, we show how fwelnet can be used in multi-task learning. We end with a discussion and ideas for future work. The appendix contains further details and proofs.

\section{Related work}\label{sec:litreview}

The idea of assigning different penalty weights for different features in the lasso or elastic net objective is not new. For example, the adaptive lasso \citep{Zou2006} assigns feature $j$ a penalty weight $w_j = 1 / |\hat{\beta}_j^{OLS}|^\gamma$, where $\hat{\beta}_j^{OLS}$ is the estimated coefficent for feature $j$ in the OLS model and $\gamma > 0$ is some hyperparameter. However, the OLS solution only depends on $\bX$ and $\by$ and does not incorporate any external information on the features. In the work closest to ours, \cite{Bergersen2011} propose using weights $w_j = \dfrac{1}{|\eta_j (\by, \bX, \bZ)|^q}$, where $\eta_j$ is some function (possibly varying for $j$) and $q$ is a hyperparameter controlling the shape of the weight function. While the authors present two ideas for what the $\eta_j$'s could be, they do not give general guidance on how to choose these functions which could drastically influence the model-fitting algorithm.

There is a correspondence between penalized regression estimates and Bayesian maximum a~posteriori (MAP) estimates with a particular choice of prior for the coefficients. (For example, ridge regression and lasso regression are MAP estimates when the coefficient vector $\beta$ is given a normal and Laplace prior respectively.) Within this Bayesian framework, some methods have been developed to use external feature information to guide the choice of prior. For example, \cite{VandeWiel2016} take an empirical Bayes approach to estimate the prior for ridge regression, while \cite{Velten2018} use variational Bayes to do so for general convex penalties.

We also note that most previous approaches for penalized regression with external information on the features only work with specific types of such information. A large number of methods have been developed to make use of \textit{feature grouping information}. Popular methods for using grouping information in penalized regression include the group lasso \citep{Yuan2006} and the overlap group lasso \citep{Jacob2009}. IPF-LASSO (integrative lasso with penalty factors) \citep{Boulesteix2017} gives features in each group its own penalty parameter, to be chosen via cross-validation. \cite{Tai2007} modify the penalized partial least squares (PLS) and nearest shrunken centroids methods to have group-specific penalties.

Other methods have been developed to incorporate ``network-like" or feature similarity information, where the user has information on how the features themselves are related to each other. For example, the fused lasso \citep{Tibshirani2005} adds an $\ell_1$ penalty on the successive differences of the coefficients to impose smoothness on the coefficient profile. Structured elastic net \citep{Slawski2010} generalizes the fused lasso by replacing the $\ell_2$-squared penalty in elastic net with $\beta^T \Lambda \beta$, where $\Lambda$ is a symmetric, positive semi-definite matrix chosen to reflect some a priori known structure between the features. \cite{Mollaysa2017} use the feature information matrix $\bZ$ to compute a feature similarity matrix, which is used to construct a regularization term in the loss criterion to be minimized. The regularizer encourages the model to give the same output as long as the total contribution of similar features is the same. We note that this approach implicitly assumes that the sources of feature information are equally relevant, which may or may not be the case.

It is not clear how most of the methods in the previous two paragraphs can be generalized to more general sources of feature information. Our method has the distinction of being able to work directly with real-valued feature information and to integrate multiple sources of feature information. We note that while \cite{VandeWiel2016} claim to be able to handle binary, nominal, ordinal and continuous feature information, the method actually ranks and groups features based on such information and only uses this grouping information in the estimation of the group-specific penalties. Nevertheless, it is able to incorporate more than one source of feature information.

\section{Feature-weighted elastic net (``fwelnet")}\label{sec:fwelnet}

As mentioned in the introduction, one direct way to utilize the scores $\bz_j^T \theta$ in model-fitting is to give each feature a different penalty weight in the elastic net objective function based on its score:

\begin{equation}
J_{\lambda, \alpha, \theta}(\beta_0,\beta) = \frac{1}{2}\|\by-\beta_0\bone -\bX \beta\|_2^2+ \lambda \sum_{j=1}^p w_j(\theta) \left[\alpha |\beta_j | + \frac{1 - \alpha}{2} \beta_j^2 \right],
\end{equation}

where $w_j(\theta) = f \left(\bz_j^T \theta \right)$ for some function $f$. Our proposed method, which we call the {\em feature-weighted elastic net}  (``fwelnet''), specifies $f$:

\begin{equation}\label{eqn:penfactors}
w_j(\theta) = \frac{\sum_{\ell = 1}^p \exp \left(\bz_\ell^T \theta \right)}{p \exp \left(\bz_j^T \theta \right)}.
\end{equation}

The fwelnet algorithm seeks the minimizer of this objective function over $\beta_0$ and $\beta$:

\begin{align}
( \hat{\beta}_0, \hat{\beta}) &= \underset{\beta_0, \beta}{\text{argmin}} \:\: J_{\lambda, \alpha, \theta}(\beta_0,\beta) \nonumber \\
&= \underset{\beta_0, \beta}{\text{argmin}} \:\: \frac{1}{2}\|\by-\beta_0\bone -\bX \beta\|_2^2+ \lambda \sum_{j=1}^p w_j(\theta) \left[\alpha |\beta_j | + \frac{1 - \alpha}{2} \beta_j^2 \right]. \label{eqn:fwelnet}
\end{align}

There are a number of reasons for this choice of penalty factors. First, when $\theta = 0$, we have $w_j(\theta) = 1$ for all $j$, reducing fwelnet to the original elastic net algorithm. Second, $w_j(\theta) \geq 1/p$ for all $j$ and $\theta$, ensuring that we do not end up with features having negligible penalty. This allows the fwelnet solution to have a wider range of sparsity as we go down the path of $\lambda$ values. Third, this formulation provides a connection between fwelnet and the group lasso \citep{Yuan2006} which we detail in Section \ref{sec:glasso}. Finally, we have a natural interpretation of a feature's score: if $\bz_j^T \theta$ is relatively large, then $w_j$ is relatively small, meaning that feature $j$ is more important for the response and hence should have smaller regularization penalty.

\begin{figure}[ht]
\centerline{\includegraphics[width=4in]{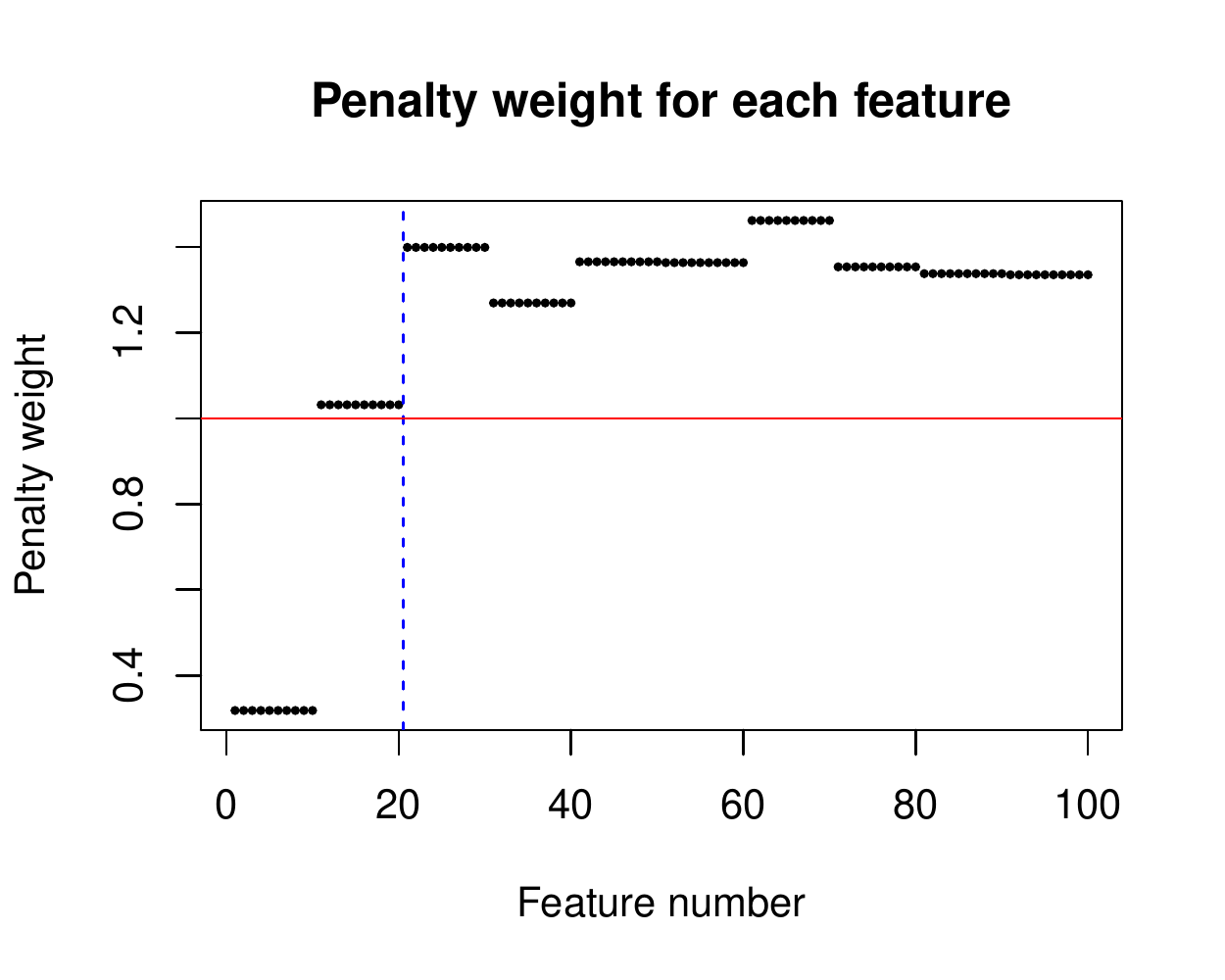}}
\caption[fig:pen_factors]{\em Penalty factors which fwelnet assigns to each feature. $n = 200$, $p = 100$ with features in groups of size $10$. The response is a noisy linear combination of the first two groups, with signal in the first group being stronger than that in the second. As expected, fwelnet's penalty weights for the true features (left of blue dotted line) are lower than that for null features. In elastic net, all features would be assigned a penalty factor of 1 (horizontal red line).}
\label{fig:pen_factors}
\end{figure}

We illustrate the last property via a simulated example. In this simulation, we have $n = 200$ observations and $p = 100$ features which come in groups of $10$. The response is a linear combination of the first two groups with additive Gaussian noise. The coefficient for the first group is $4$ while the coefficient for the second group is $-2$ so that the features in the first group exhibit stronger correlation to the response compared to the second group. The ``features of features" matrix $\bZ \in \bbR^{100 \times 10}$ is grouping information, i.e. $z_{jk} = 1$ if feature $j$ belongs to group $k$, and is $0$ otherwise. Figure \ref{fig:pen_factors} shows the penalty factors $w_j$ that fwelnet assigns the features. As one would expect, the features in the first group have the smallest penalty factor followed by features in the second group. In contrast, the original elastic net algorithm would assign penalty factors $w_j = 1$ for all $j$.

\subsection{Computing the fwelnet solution}\label{sec:computation}

It can be easily shown that $\hat{\beta}_0 = \overline{\by} - \sum_{j=1}^p \hat{\beta}_j \overline{\bx}_{\cdot j}$. Henceforth, we assume that $\by$ and the columns of $\bX$ are centered so that $\hat{\beta}_0 = 0$ and we can ignore the intercept term in the rest of the discussion.

For given values of $\lambda$, $\alpha$ and $\theta$, it is easy to solve \eqref{eqn:fwelnet}: the objective function is convex in $\beta$ (in fact it is piecewise-quadratic in $\beta$) and $\hat{\beta}$ can be found efficiently using algorithms such as coordinate descent. However, to deploy fwelnet in practice we need to determine the hyperparameter values $\hat{\lambda} \in \bbR$, $\hat{\alpha} \in \bbR$ and $\hat{\theta} \in \bbR^K$ that give good performance. When $K$, the number of sources of feature information, is small, one could run the algorithm for a grid of $\theta$ values, then pick the value which gives the smallest cross-validated loss. Unfortunately, this approach is computationally infeasible for even moderate values of $K$.

To avoid this computational bottleneck, we propose Algorithm \ref{alg_fwelnet} as a method to find $\hat{\beta}$ and $\hat{\theta}$ at the same time. If we think of $\theta$ as an argument of the objective function $J$, Step 3 can be thought of as alternating minimization over $\theta$ and $\beta$. Notice that in Step 3(c), we allow the algorithm to have a different value of $\hat{\beta}$ for each $\lambda$ value. However, {\em we force $\hat{\theta}$ to be the same across all $\lambda$ values}: Steps (a) and (b) can be thought of as a heuristic to perform gradient descent for $\theta$ under this constraint.

\begin{algorithm}
\caption{ \em Fwelnet algorithm}
\label{alg_fwelnet}
\begin{enumerate}
\item Select a value of $\alpha \in [0, 1]$ and a sequence of $\lambda$ values $\lambda_1 > \ldots > \lambda_m$.

\item For $i = 1, \ldots, m$, initialize $\beta^{(0)}(\lambda_i)$ at the elastic net solution for the corresponding $\lambda_i$. Initialize $\theta^{(0)} = \bzero$.

\item For $k = 0, 1, \ldots$ until convergence:

		\begin{enumerate}
		\item Set $\Delta \theta$ to be the component-wise mean/median of $\left. \dfrac{\partial J_{\lambda_i, \alpha}}{\partial \theta} \right|_{\beta = \beta^{(k)}, \theta = \theta^{(k)}}$ over $i = 1, \dots, m$.
		
		\item Set $\theta^{(k+1)} = \theta^{(k)} - \eta \Delta \theta$, where $\eta$ is the step size computed via backtracking line search to ensure that the mean/median of $J_{\lambda_i, \alpha} \left(\beta^{(k)}, \theta^{(k+1)} \right)$ over $i = 1, \ldots, m$ is less than that for $J_{\lambda_i, \alpha} \left(\beta^{(k)}, \theta^{(k)} \right)$.
		
		\item For $i = 1, \ldots, m$, set $\beta^{(k+1)}(\lambda_i) = $ elastic net solution for $\lambda_i$ where the penalty factor for feature $j$ is $w_j (\theta^{(k+1)})$.
		\end{enumerate}

\end{enumerate}

\end{algorithm}

We have developed an R package, \texttt{fwelnet}, which implements Algorithm \ref{alg_fwelnet}. We note that Step 3(c) of Algorithm \ref{alg_fwelnet} can be done easily by using the \texttt{glmnet} function in the \texttt{glmnet} R package and specifying the \texttt{penalty.factor} option. In practice, we use the lambda sequence $\lambda_1 > \dots > \lambda_m$ provided by \texttt{glmnet}'s implementation of the elastic net as this range of $\lambda$ values covers a sufficiently wide range of models. With this choice of $\lambda$ sequence, we find that fwelnet's performance does not change much whether we use the component-wise mean or median in Step 3(a), or the mean or median in Step 3(b). Also, instead of running Step 3 until convergence, we recommend running it for a small fixed number of iterations $N$. Step 3(c) is the bottleneck of the algorithm, and so the runtime for fwelnet is approximately $N+1$ times that of \texttt{glmnet}. In our simulation studies, we often find that one pass of Step 3 gives a sufficiently good solution. We suggest treating $N$ as a hyperparameter and running fwelnet for $N = 1, 2$ and $5$.

(We also considered an approach where we did not constrain the value of $\theta$ to be equal across $\lambda$ values. While conceptually straightforward, the algorithm was computationally slow and did not perform as well as Algorithm \ref{alg_fwelnet} in prediction. A sketch of this approach can be found in Appendix \ref{sec:theta_parameter}.)

\subsection{Extending fwelnet to generalized linear models (GLMs)}

In the exposition above, the elastic net is described as a regularized version of the ordinary least squares model. It is easy to extend elastic net regularization to generalized linear models (GLMs) by replacing the RSS term with the negative log-likelihood of the data:

\begin{equation}\label{eqn:elnet_glm}
(\hat{\beta}_0, \hat{\beta}) = \underset{\beta_0, \beta}{\text{argmin}} \:\: \sum_{i=1}^n \ell \left( y_i, \beta_0 + \sum_{j=1}^p x_{ij} \beta_j \right) + \lambda \sum_{j=1}^p \left[\alpha |\beta_j | + \frac{1 - \alpha}{2} \beta_j^2 \right],
\end{equation}

where $\ell(y_i, \beta_0 + \sum_j x_{ij}\beta_j)$ is the negative log-likelihood contribution of observation $i$. Fwelnet can be extended to GLMs in a similar fashion:

\begin{equation}\label{eqn:fwelnet_glm}
(\hat{\beta}_0, \hat{\beta}, \hat{\theta}) = \underset{\beta_0, \beta, \theta}{\text{argmin}} \:\: \sum_{i=1}^n \ell \left( y_i, \beta_0 + \sum_{j=1}^p x_{ij} \beta_j \right) + \lambda \sum_{j=1}^p w_j(\theta) \left[\alpha |\beta_j | + \frac{1 - \alpha}{2} \beta_j^2 \right],
\end{equation}

with $w_j(\theta)$ as defined in \eqref{eqn:penfactors}. Theoretically Algorithm \ref{alg_fwelnet} can be used as-is to solve \eqref{eqn:fwelnet_glm}. Because $\theta$ only appears in the penalty term and not in the negative log-likelihood, this extension is not hard to implement in code. The biggest hurdle to this extension is a solver for \eqref{eqn:elnet_glm} which is needed for Steps 2 and 3(c). Step 3(a) is the same as before, while Step 3(b) simply requires a function that allows us to compute the negative log-likelihood $\ell$.

\section{A simulation study}\label{sec:sim}
 
We tested the performance of fwelnet against other methods in a simulation study. In the three settings studied, the true signal is a linear combination of the columns of $\bX$, with the true coefficient vector $\beta$ being sparse. The response $\by$ is the signal corrupted by additive Gaussian noise. In each setting, we gave different types of feature information to fwelnet to determine the method's effectiveness.

For all methods, we used cross-validation (CV) to select the tuning parameter $\lambda$. Unless otherwise stated, the $\alpha$ hyperparameter was set to $1$ (i.e. no $\ell_2$ squared penalty) and Step 3 of Algorithm \ref{alg_fwelnet} was run for one iteration, with the mean used for Steps 3(a) and 3(b). To compare methods, we considered the mean squared error (MSE) $MSE = \bbE [(\hat{y} - \mu)^2]$ achieved on 10,000 test points, as well as the true positive rate (TPR) and false positive rate (FPR) of the fitted models. (The oracle model which knows the true coefficient vector $\beta$ has a test MSE of $0$.) We ran each simulation 30 times to get estimates for these quantities. (See Appendix \ref{sec:simdetails} for details of the simulations.)

\subsection{Setting 1: Noisy version of the true $|\beta|$}

In this setting, we have $n = 100$ observations and $p = 50$ features, with the true signal being a linear combination of just the first 10 features. The feature information matrix $\bZ$ has two columns: a noisy version of $|\beta|$ and a column of ones.

\begin{figure}[!htpb]
\centerline{\includegraphics[width=3in,valign=t]{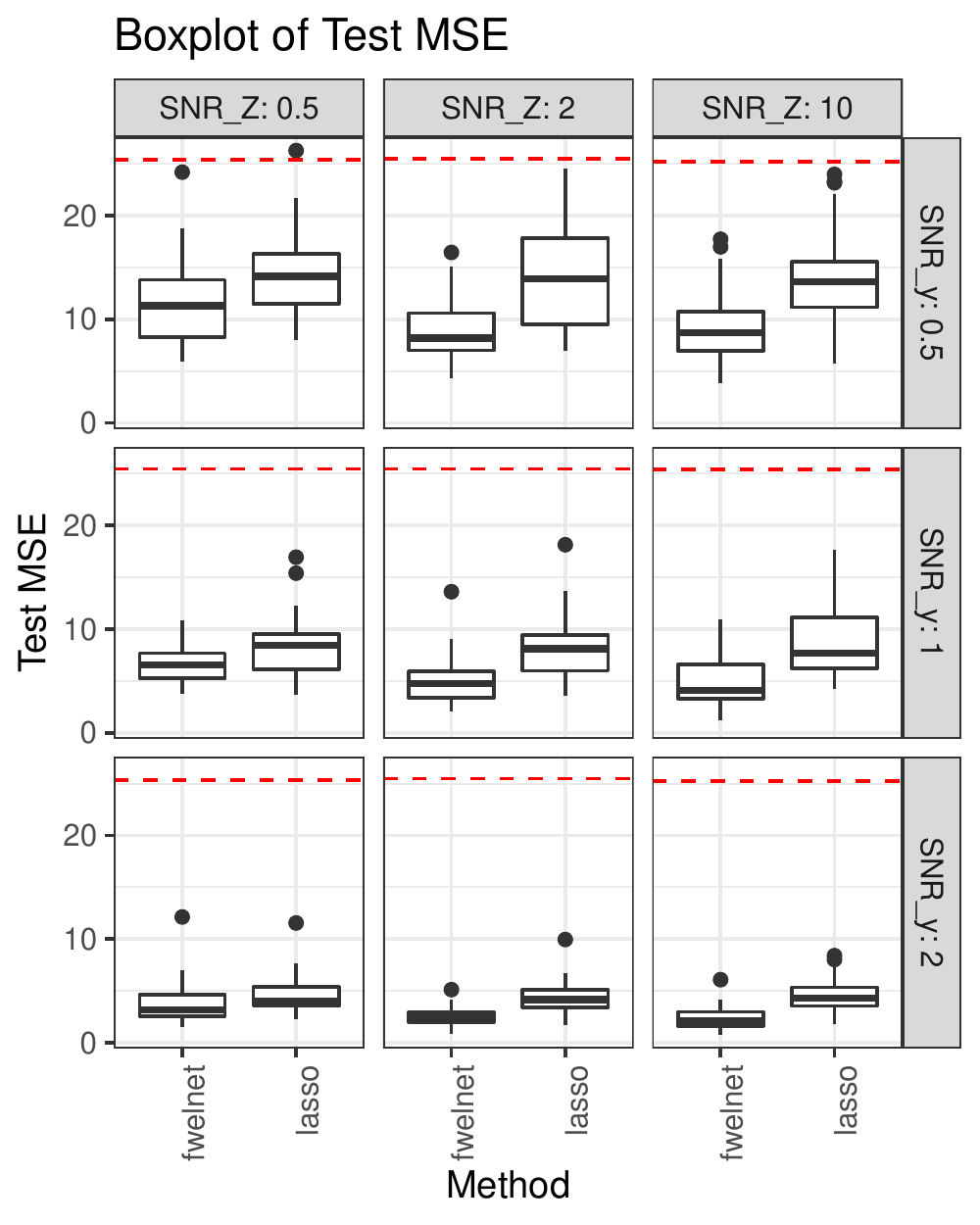}\includegraphics[width=3in,valign=t]{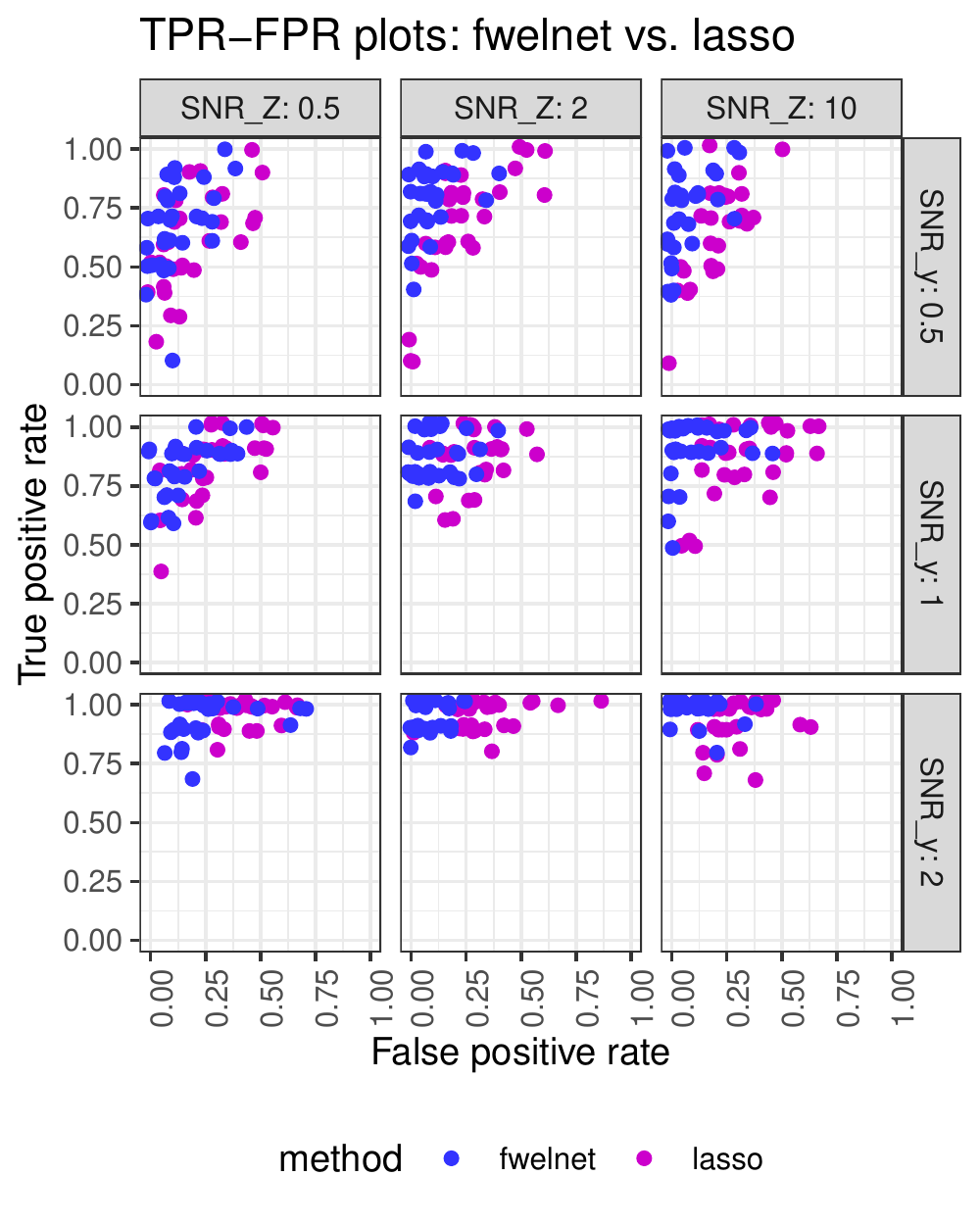}}
\caption[fig:sim1]{\em ``Feature of features": noisy version of the true $|\beta|$. $n = 100$, $p = 50$. The response is a linear combination of the first 10 features. As we go from left to right, the signal-to-noise ratio (SNR) for $\by$ increases; as we go from top to bottom, the SNR in $\bZ$ increases. The panel on the left shows the raw test mean squared error (MSE) figures with the red dotted line indicating the median null test MSE. In the figure on the right, each point depicts the true positive rate (TPR) and false positive rate (FPR) of the fitted model for one of 30 simulation runs. Fwelnet outperforms the lasso in test MSE, with the improvement getting larger as the SNR in $\bZ$ increases. Fwelnet appears to have similar TPR to the lasso but has significantly smaller FPR.}
\label{fig:sim1}
\end{figure}

We compared fwelnet against the lasso (using the \texttt{glmnet} package) across a range of signal-to-noise ratios (SNR) in both the response $\by$ and the feature information matrix $\bZ$ (see details in Appendix \ref{sec:simdetails_1}). The results are shown in Figure \ref{fig:sim1}. As we would expect, the test MSE figures for both methods decreased as the SNR in the response increased. The improvement of fwelnet over the lasso increased as the SNR in $\bZ$ increased. In terms of feature selection, fwelnet appeared to have similar TPR as the lasso but had smaller FPR.

\subsection{Setting 2: Grouped data setting}

In this setting, we have $n = 100$ observations and $p = 150$ features, with the features coming in 15 groups of size 10. The feature information matrix $\bZ \in \bbR^{150 \times 15}$ contains group membership information for the features: $z_{jk} = 1 \{ \text{feature } j \in \text{group } k \}$. We compared fwelnet against the lasso and the group lasso (using the \texttt{grpreg} package) across a range of signal-to-noise ratios (SNR) in the response $\by$.

\begin{figure}[!htpb]
\centerline{\includegraphics[width=3in,valign=t]{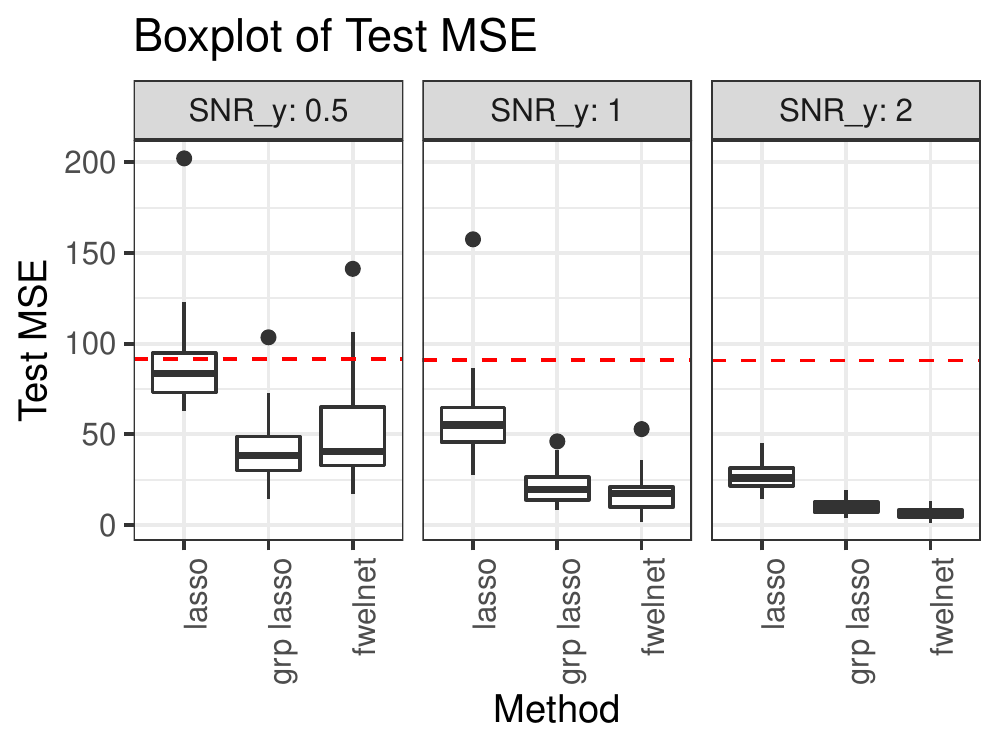}\includegraphics[width=3in,valign=t]{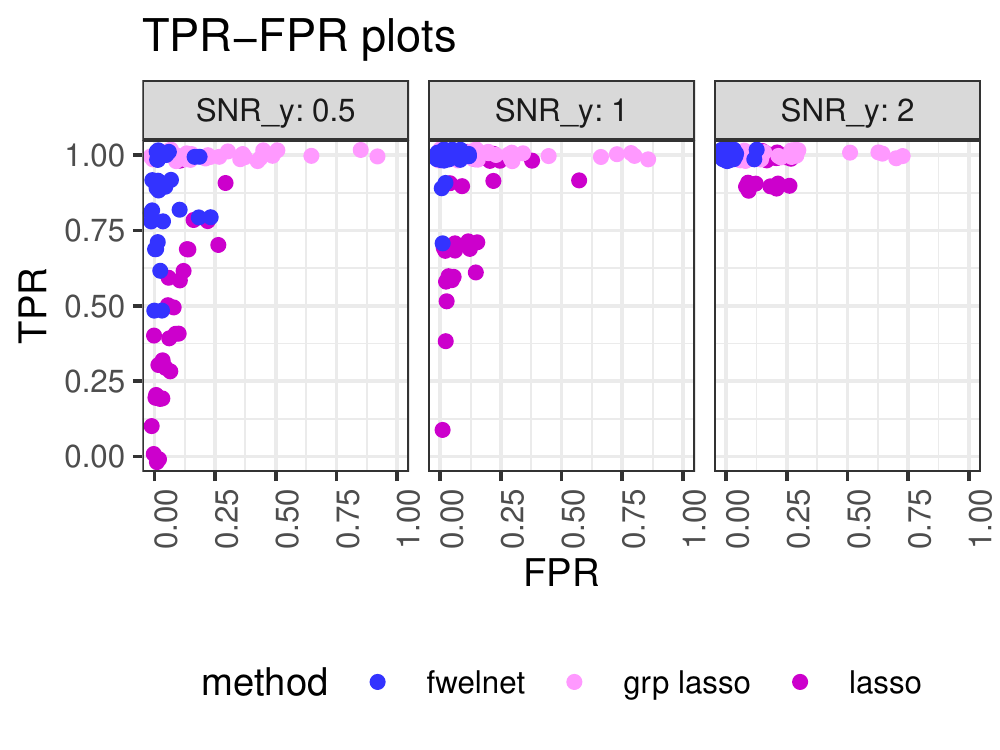}}
\caption[fig:sim2]{\em ``Feature of features": grouping data. $n = 100$, $p = 150$. The features come in groups of 10, with the response being a linear combination of the features in the first group. As we go from left to right, the signal-to-noise ratio (SNR) for $\by$ increases. The figure on the left shows the test mean squared error (MSE) results with the red dotted line indicating the median null test MSE. In the figure on the right, each point depicts the true positive rate (TPR) and false positive rate (FPR) of the fitted model for one of 30 simulation runs. Fwelnet outperforms the group lasso in terms of test MSE at higher SNR levels. Fwelnet has higher TPR than the lasso, and lower FPR than the group lasso.}
\label{fig:sim2}
\end{figure}

We considered two different responses in this setting. The first response we considered was a linear combination of the features in the first group only, with additive Gaussian noise. The results are depicted in Figure \ref{fig:sim2}. In terms of test MSE, fwelnet was competitive with the group lasso in the low SNR scenario and came out on top for the higher SNR settings. In terms of feature selection, fwelnet had comparable TPR as the group lasso but drastically smaller FPR. Fwelnet had better TPR and FPR than the lasso in this case. We believe that fwelnet's improvement over the group lasso could be because the true signal was sparse: fwelnet's connection to the $\ell_1$ version of the group lasso (see Section \ref{sec:glasso} for details) encourages greater sparsity than the usual group lasso penalty based on $\ell_2$ norms.

\begin{figure}[!htpb]
\centerline{\includegraphics[width=3in,valign=t]{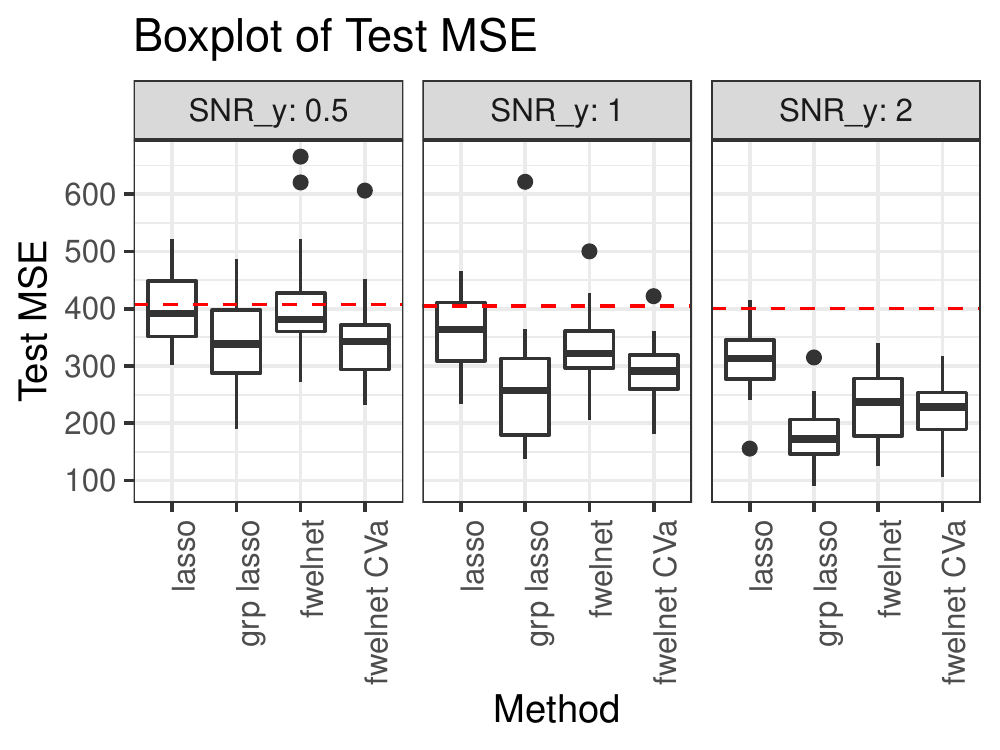}\includegraphics[width=3in,valign=t]{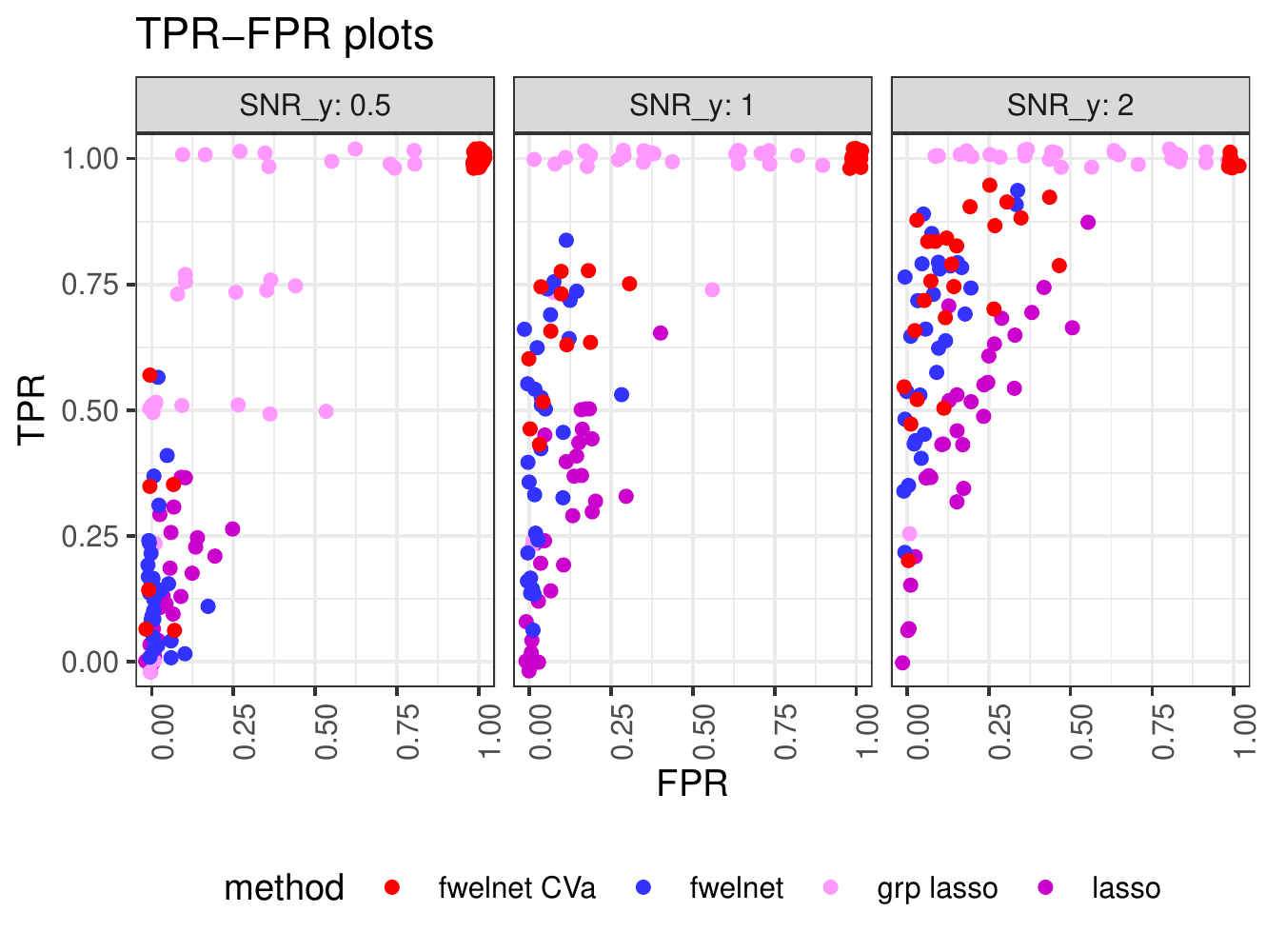}}
\caption[fig:sim2_4groups]{\em ``Feature of features": grouping data. $n = 100$, $p = 150$. The features come in groups of 10, with the response being a linear combination of the first 4 groups. As we go from left to right, the signal-to-noise ratio (SNR) for $\by$ increases. The figure on the left shows the test mean squared error (MSE) results with the red dotted line indicating the median null test MSE. Fwelnet sets $\alpha = 1$ while fwelnet CVa selects $\alpha$ via cross-validation. In the figure on the right, each point depicts the true positive rate (TPR) and false positive rate (FPR) of the fitted model for one of 30 simulation runs. Group lasso performs best here. Cross-validation for $\alpha$ improves test MSE performance but at the expense of having very high FPR.}
\label{fig:sim2_4groups}
\end{figure}

The second response we considered in this setting was not as sparse in the features: the true signal was a linear combination of the first 4 feature groups. The results are shown in Figure \ref{fig:sim2_4groups}. In this case, the group lasso performed better than fwelnet when the hyperparameter $\alpha$ was fixed at 1, which is in line with our earlier intuition that fwelnet would perform better in sparser settings. It is worth noting that fwelnet with $\alpha = 1$ performs appreciably better than the lasso when the SNR is higher. Selecting $\alpha$ via cross-validation improved the test MSE performance of fwelnet, but not enough to outperform the group lasso. The improvement in test MSE also came at the expense of very high FPR.

\subsection{Setting 3: Noise variables}

In this setting, we have $n = 100$ observations and $p = 100$ features, with the true signal being a linear combination of just the first 10 features. The feature information matrix $\bZ$ consists of 10 noise variables that have nothing to do with the response. Since fwelnet is adapting to these features, we expect it to perform worse than comparable methods.

\begin{figure}[!htpb]
\centerline{\includegraphics[width=3in,valign=t]{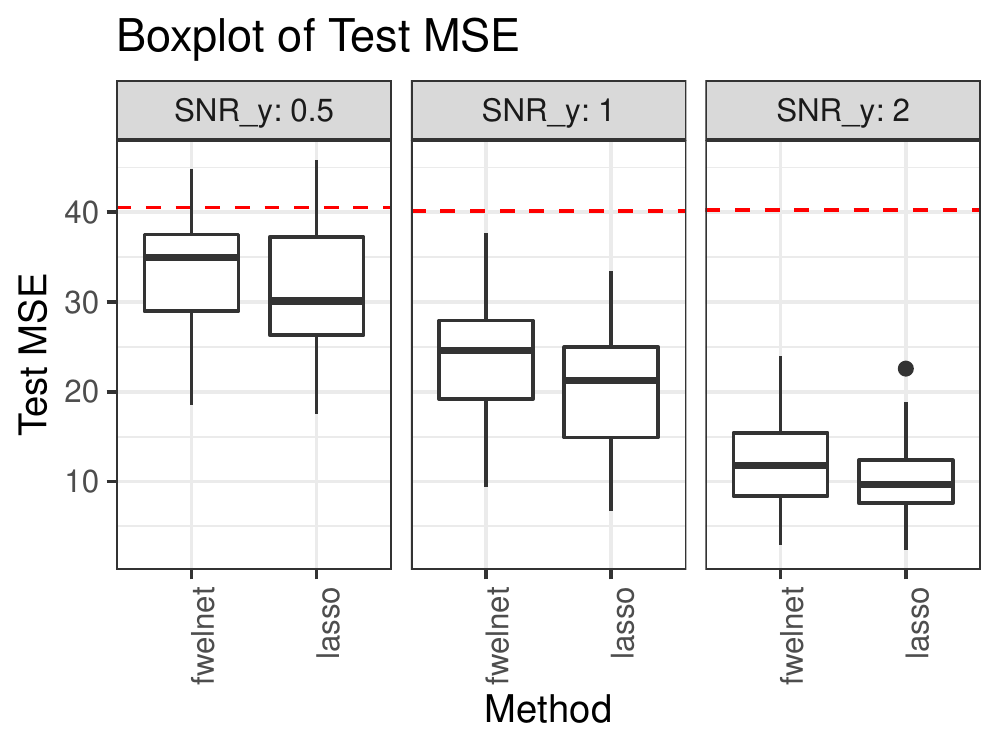}\includegraphics[width=3in,valign=t]{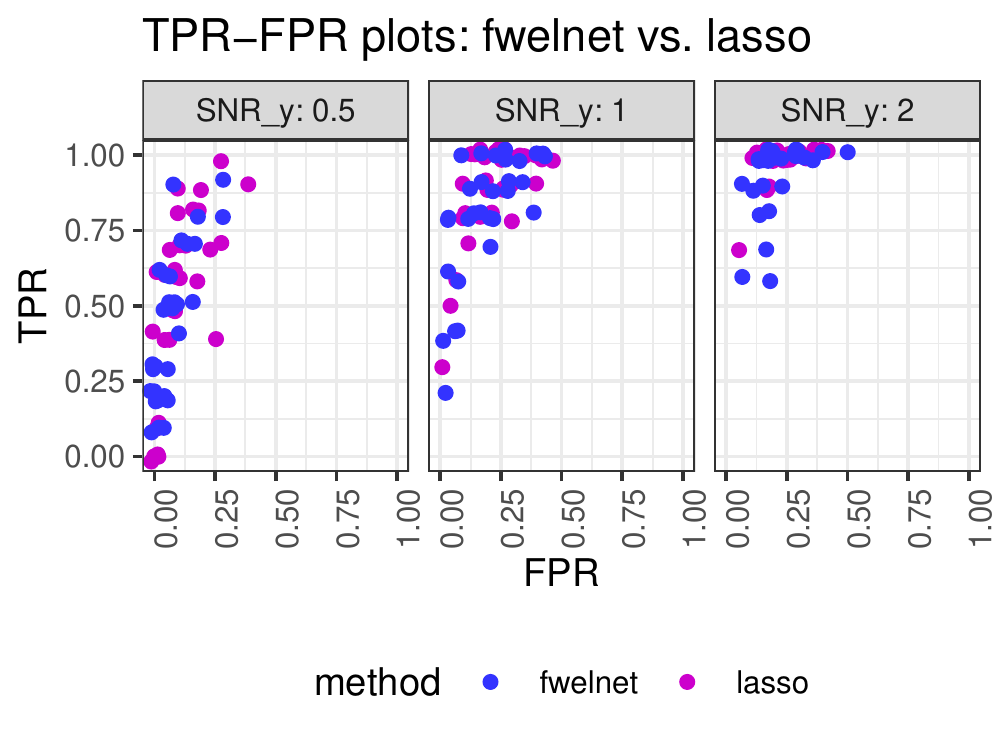}}
\caption[fig:sim3]{\em ``Feature of features": 10 noise variables. $n = 100$, $p = 100$. The response is a linear combination of the first 10 features. As we go from left to right, the signal-to-noise ratio (SNR) for $\by$ increases. The figure on the left shows the test mean squared error (MSE) results, with the red dotted line indicating the median null test MSE. In the figure on the right, each point depicts the true positive rate (TPR) and false positive rate (FPR) of the fitted model for one of 30 simulation runs. Fwelnet only performs slightly worse than the lasso in test MSE, and has similar TPR and FPR as the lasso.}
\label{fig:sim3}
\end{figure}

We compare fwelnet against the lasso across a range of signal-to-noise ratios (SNR) in the response $\by$. The results are depicted in Figure \ref{fig:sim3}. As expected, fwelnet has higher test MSE than the lasso, but the decrease in performance is not drastic. Fwelnet attained similar FPR and TPR to the lasso.

\section{Application: Early prediction of preeclampsia}\label{sec:realdata}

Preeclampsia is a leading cause of maternal and neonatal morbidity and mortality, affecting 5 to 10 percent of all pregnancies. The biological and phenotypical signals associated with late-onset preeclampsia strengthen during the course of pregnancy, often resulting in a clinical diagnosis after 20 weeks of gestation \citep{Zeisler2016}. An earlier test for prediction of late-onset preeclampsia will enable timely interventions for improvement of maternal and neonatal outcomes \citep{Jabeen2011}. In this example, we seek to leverage data collected in late pregnancy to guide the optimization of a predictive model for early diagnosis of late-onset preeclampsia.

We used a dataset of plasma proteins measured during various gestational ages of pregnancy \citep{Erez2017}. For this example, we considered time points $\leq 20$ weeks ``early" and time points $> 20$ weeks as ``late". We had measurements for between 2 to 6 time points for each of the 166 patients for a total of 666 time point observations. Protein measurements were log-transformed to reduce skew. We first split the patients equally into two buckets. For patients in the first bucket, we used only their late time points (83 patients with 219 time points) to train an elastic net model with $\alpha = 0.5$ to predict whether the patient would have preeclampsia. From this late time point model, we extracted model coefficients at the $\lambda$ hyperparameter value which gave the highest 10-fold cross-validated (CV) area under the curve (AUC). For patients in the second bucket, we used only their early time points (83 patients with 116 time points) to train a fwelnet model with the absolute values of the late time point model coefficients as feature information. When performing CV, we made sure that observations from one patient all belonged to the same CV fold to avoid ``contamination" of the held-out fold. One can also run the fwelnet model with additional sources of feature information for each of the proteins.

\begin{figure}[!htpb]
\centerline{\includegraphics[width=4in]{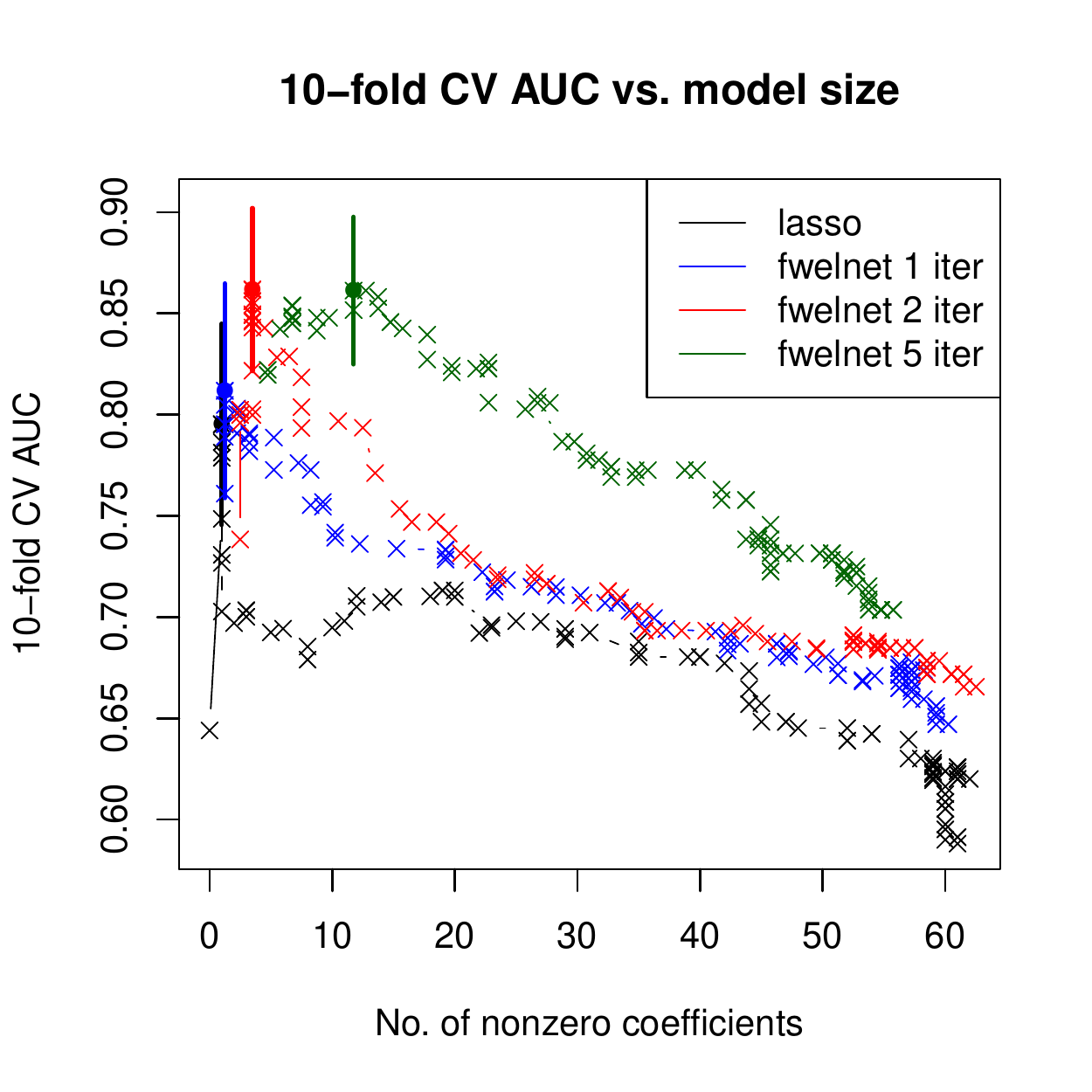}}
\caption[fig:pe]{\em Early prediction of pre-eclampsia: Plot of 10-fold cross-validated (CV) area under the curve (AUC). 10-fold CV AUC is plotted against the number of non-zero coefficients for each model trained on just early time point data. The baseline method is the lasso; we run fwelnet for 1, 2 and 5 minimization iterations. For each method/algorithm, the model with highest CV AUC is marked by a dot. To reduce clutter in the figure, the $\pm 1$ standard error bars are drawn for just these models. Fwelnet with 2 minimization iterations has the largest CV AUC.}
\label{fig:pe}
\end{figure}

We compare the 10-fold CV AUC for fwelnet run with 1, 2 and 5 minimization iterations (i.e. Step 3 in Algorithm \ref{alg_fwelnet}) against the lasso as a baseline. Figure \ref{fig:pe} shows a plot of 10-fold CV AUC for these methods against the number of features with non-zero coefficients in the model. The lasso obtains a maximum CV AUC of 0.80, while fwelnet with 2 minimization iterations obtains the largest CV AUC of 0.86 among all methods.

We note that the results were somewhat dependent on (i) how the patients were split into the early and late time point models, and (ii) how patients were split into CV folds when training each of these models. We found that if the late time point model had few non-zero coefficients, then the fwelnet model for the early time point data was very similar to that for the lasso. This matches our intuition: if there are few non-zero coefficients, then we are injecting very little additional information through the relative penalty factors in fwelnet, and so it will give a very similar model to elastic net. Nevertheless, we did not encounter cases where running fwelnet gave worse CV AUC than the lasso.

\section{Connection to the group lasso}\label{sec:glasso}

One common setting where ``features of features" arise naturally is when the features come in non-overlapping groups. Assume that the features in $\bX$ come in $K$ non-overlapping groups. Let $p_k$ denote the number of features in group $k$, and let $\beta^{(k)}$ denote the subvector of $\beta$ which belongs to group $k$. Assume also that $\by$ and the columns of $\bX$ are centered so that $\hat{\beta}_0 = 0$. In this setting, \cite{Yuan2006} introduced the group lasso estimate as the solution to the optimization problem

\begin{equation}\label{eqn:glasso1}
\underset{\beta}{\minimize} \quad \frac{1}{2} \ltwo{\by - \bX \beta}^2 + \lambda \sum_{k=1}^K \ltwo{\beta^{(k)}}.
\end{equation}

The $\ell_2$ penalty on features at the group level ensures that features belonging to the same group are either all included in the model or all excluded from it. Often, the penalty given to group $k$ is modified by a factor of $\sqrt{p_k}$ to take into account the varying group sizes:
\begin{equation}\label{eqn:glasso2}
\hat{\beta}_{gl, 2} (\lambda) = \underset{\beta}{\text{argmin}} \quad \frac{1}{2} \ltwo{\by - \bX \beta}^2 + \lambda \sum_{k=1}^K \sqrt{p_k} \ltwo{\beta^{(k)}}.
\end{equation}

Theorem \ref{thm:glasso} below establishes a connection between fwelnet and the group lasso. For the moment, consider the more general penalty factor $w_j(\theta) = \dfrac{\sum_{\ell = 1}^p f(\bz_\ell^T \theta)}{pf(\bz_j^T \theta)}$, where $f$ is some function with range $[0, +\infty)$. (Fwelnet makes the choice $f(x) = e^x$.)

\begin{theorem}\label{thm:glasso}
If the ``features of features" matrix $\bZ \in \bbR^{p \times K}$ is given by $z_{jk} = 1\{ \text{feature } j \in \text{ group } k \}$, then minimizing the fwelnet objective function \eqref{eqn:fwelnet} jointly over $\beta_0$, $\beta$ and $\theta$ reduces to

\begin{align*}
&\underset{\beta}{\text{argmin}} \quad \frac{1}{2} \ltwo{\by - \bX \beta}^2 + \lambda' \sum_{k = 1}^K \sqrt{p_k \left[ \alpha \lone{\beta^{(k)}} + \frac{1-\alpha}{2} \ltwo{\beta^{(k)}}^2 \right]}  \\ 
&= \begin{cases} \underset{\beta}{\text{argmin}} \quad \frac{1}{2} \ltwo{\by - \bX \beta}^2 + \lambda' \sum_{k = 1}^K \sqrt{p_k}\ltwo{\beta^{(k)}}  &\text{if } \alpha = 0, \\
\underset{\beta}{\text{argmin}} \quad  \frac{1}{2} \ltwo{\by - \bX \beta}^2 + \lambda' \sum_{k = 1}^K \sqrt{p_k \lone{\beta^{(k)}}} &\text{if } \alpha = 1, \end{cases}
\end{align*}
for some $\lambda' \geq 0$.
\end{theorem}

We recognize the $\alpha = 0$ case as minimizing the residual sum of squares (RSS) and the group lasso penalty, while the $\alpha = 1$ case is minimizing the RSS and the $\ell_1$ version of the group lasso penalty. The proof of Theorem \ref{thm:glasso} can be found in Appendix \ref{sec:glasso_proof}.

\section{Using fwelnet for multi-task learning}\label{sec:multitask}

We turn now to an application of fwelnet to \textit{multi-task learning}. In some applications, we have a single model matrix $\bX$ but are interested in multiple responses $\by_1, \dots \by_B$. If there is some common structure between the signals in the $B$ responses, it can be advantageous to fit models for them simultaneously. This is especially the case if the signal-to-noise ratios in the responses are low.

We demonstrate how fwelnet can be used to learn better models in the setting with two responses, $\by_1$ and $\by_2$. The idea is to use the absolute value of coefficients of one response as the external information for the other response. That way, a feature which has larger influence on one response is likely to be given a correspondingly lower penalty weight when fitting the other response. Algorithm \ref{alg_multitask} presents one possible way of doing so.

\begin{algorithm}
\caption{ \em Using fwelnet for multi-task learning}
\label{alg_multitask}
\begin{enumerate}
\item Initialize $\beta_1^{(0)}$ and $\beta_2^{(0)}$ at the \texttt{lambda.min} elastic net solutions for $(\bX, \by_1)$ and $(\bX, \by_2)$ respectively, that is, the value of the hyperparameter $\lambda$ which minimizes cross-validated error.

\item For $k = 0, 1, \dots$ until convergence:
		\begin{enumerate}
		\item Set $\bZ_2 = \left| \beta_1^{(k)} \right|$. Run fwelnet with $(\bX, \by_2, \bZ_2)$ and set $\beta_2^{(k+1)}$ to be the \texttt{lambda.min} solution.
		
		\item Set $\bZ_1 = \left| \beta_2^{(k+1)} \right|$. Run fwelnet with $(\bX, \by_1, \bZ_1)$ and set $\beta_1^{(k+1)}$ to be the \texttt{lambda.min} solution.
		\end{enumerate}

\end{enumerate}

\end{algorithm}

We tested the effectiveness of Algorithm \ref{alg_multitask} (with step 2 run for 3 iterations) on simulated data. We generate 150 observations with 50 independent features. The signal in response 1 is a linear combination of features 1 to 10, while the signal in response 2 is a linear combination of features 1 to 5 and 11 to 15. The coefficients are set such that those for the common features (i.e. features 1 to 5) have larger absolute value than those for the features specific to one response. The signal-to-noise ratios (SNRs) in response 1 and response 2 are 0.5 and 1.5 respectively. (See Appendix \ref{sec:sim_multitask} for more details of the simulation.)

We compared Algorithm \ref{alg_multitask} against: (i) the \textit{individual lasso (ind\_lasso)}, where the lasso is run separately for $\by_1$ and $\by_2$; and (ii) the \textit{multi-response lasso (mt\_lasso)}, where coefficients belonging to the same feature across the responses are given a joint $\ell_2$ penalty. Because of the $\ell_2$ penalty, a feature is either included or excluded in the model for all the responses at the same time.

\begin{figure}[!htpb]
\centerline{\includegraphics[width=3in]{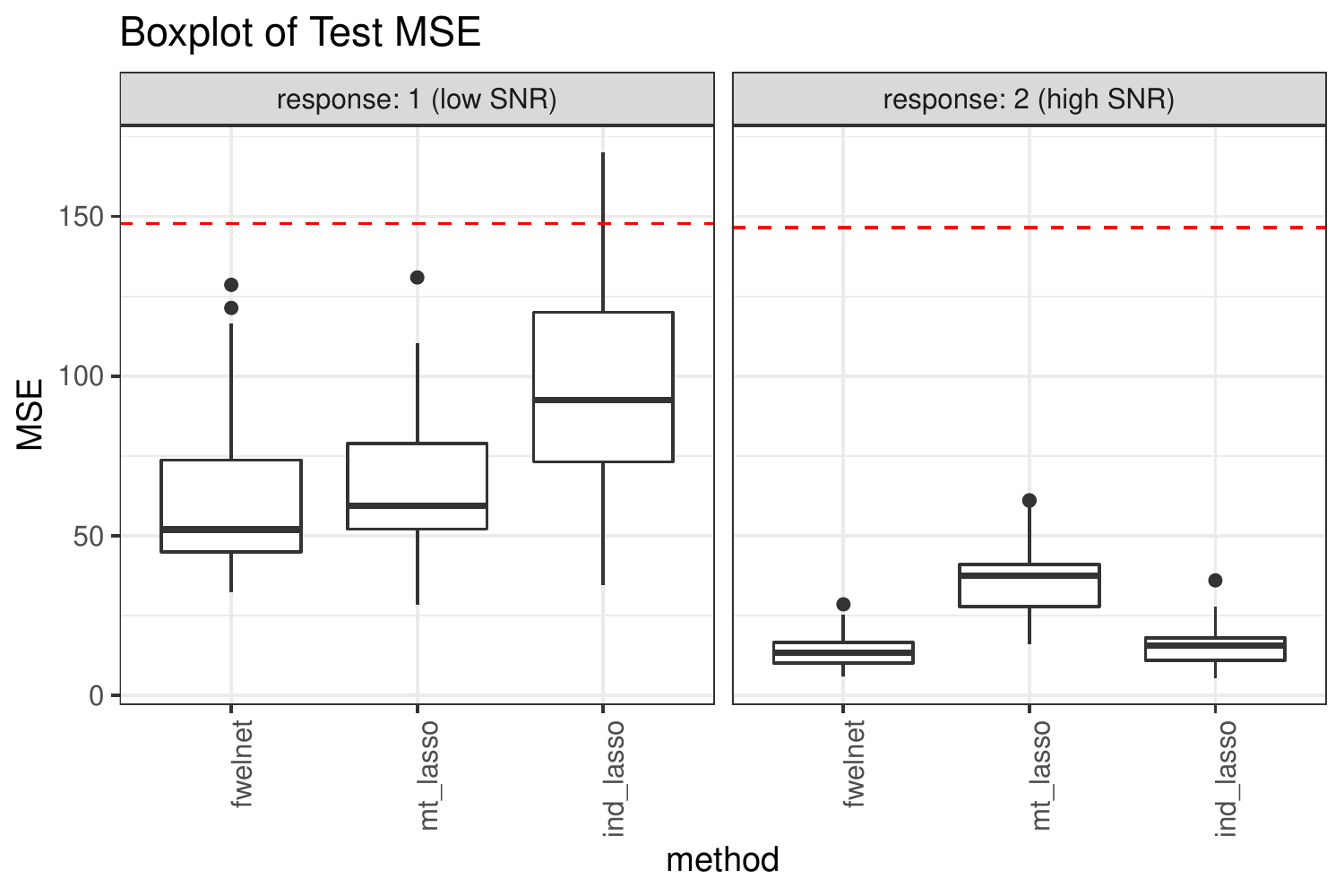} \includegraphics[width=3in]{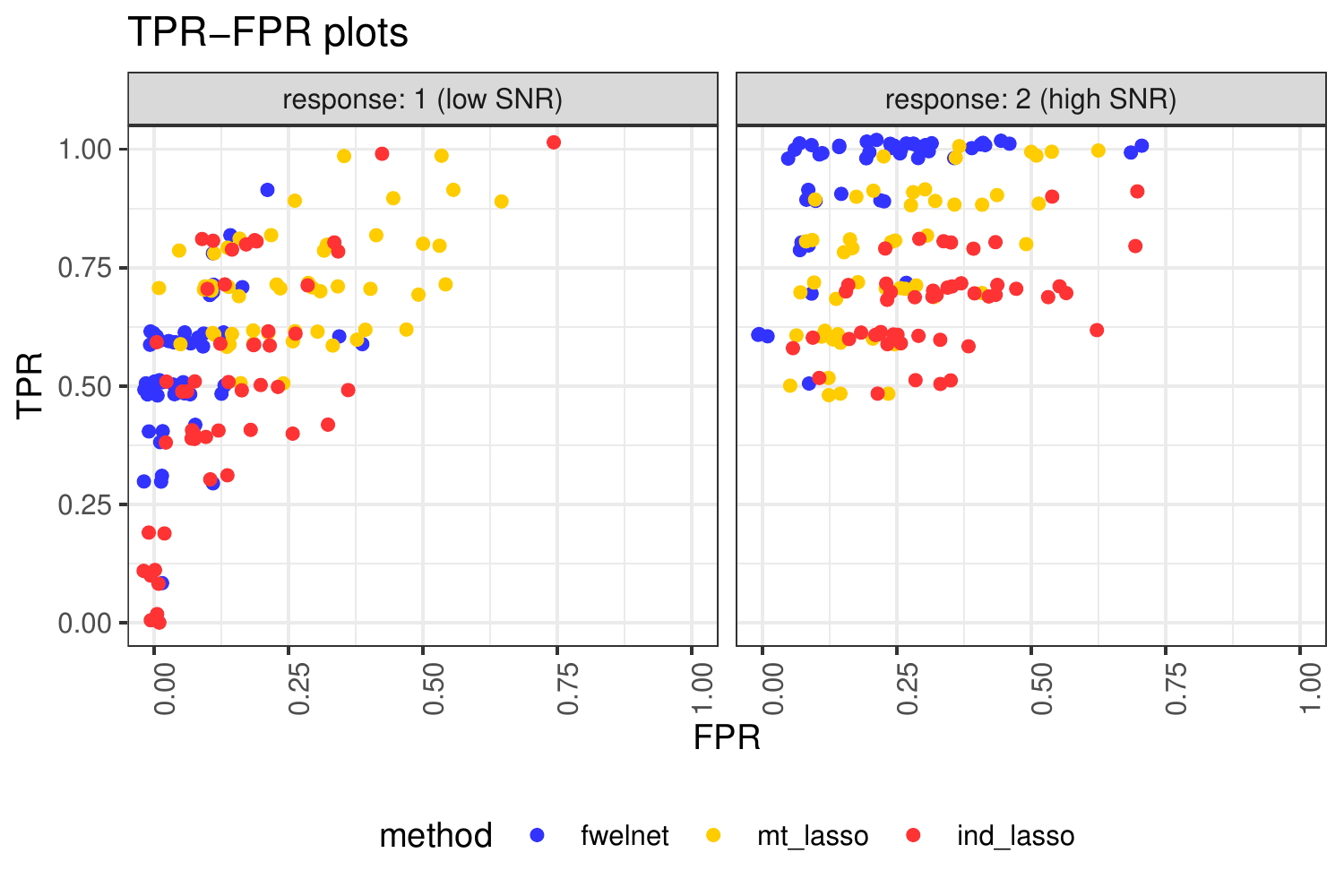}}
\caption[fig:sim_multitask]{\em Application of fwelnet to multi-task learning. $n = 150$, $p = 50$. Response 1 is a linear combination of features 1 to 10, while response 2 is a linear combination of features 1 to 5 and 11 to 15. The signal-to-noise ratios (SNR) for responses 1 and 2 are 0.5 and 1.5 respectively. The figure on the left shows the raw test mean squared error (MSE) figures with the red dotted line indicating the median null test MSE. The figure on the right shows the true positive rate (TPR) and false positive rate (FPR) of the fitted model (each point being one of 50 simulation runs). Fwelnet outperforms the individual lasso and the multi-response lasso in test MSE for both responses. Fwelnet also has better TPR and FPR than the other methods.}
\label{fig:sim_multitask}
\end{figure}

The results are shown in Figure \ref{fig:sim_multitask} for 50 simulation runs. Fwelnet outpeforms the other two methods in test MSE as evaluated on 10,000 test points. As expected, the lasso run individually for each response performs well in the response with higher SNR but poorly in the response with lower SNR. The multi-response lasso is able to borrow strength from the higher SNR response to obtain good performance on the lower SNR response. However, because the models for both responses are forced to consist of the same set of features, performance suffers on the higher SNR response. Fwelnet has the ability to borrow strength across responses without being hampered by this restriction.

\section{Discussion}\label{sec:discussion}

In this paper, we have proposed organizing information about predictor variables, which we call ``features of features", as a matrix $\bZ \in \bbR^{p \times K}$, and modifying model-fitting algorithms by assigning each feature a score, $\bz_j^T \theta$, based on this auxiliary information. We have proposed one such method, the feature-weighted elastic net (``fwelnet"), which imposes a penalty modification factor $w_j(\theta) = \dfrac{\sum_{\ell = 1}^p \exp \left(\bz_\ell^T \theta \right)}{p \exp \left(\bz_j^T \theta \right)}$ for the elastic net algorithm.

There is much scope for future work:

\begin{itemize}
\item \textit{Choice of penalty modification factor.} While the penalty modification factors $p_j(\theta)$ we have proposed works well in practice and has several desirable properties, we make no claim about its optimality. We also do not have well-developed theory for the current choice penalty factor.

\item \textit{Extending the use of scores beyond the elastic net.} The use of feature scores $\bz_j^T \theta$ in modifying the weight given to each feature in the model-fitting process is a general idea that could apply to any supervised learning algorithm. More work needs to be done on how such scores can be incorporated, with particular focus on how $\theta$ can be learned through the algorithm.

\item \textit{Whether $\theta$ should be treated as a parameter or a hyperparameter, and how to determine its value.} In this paper, we introduced $\theta$ as a hyperparameter for \eqref{eqn:fwelnet}. This formulation gives us a clear interpretation for $\theta$: $\theta_k$ is a proxy for how important the $k$th source of feature information is for identifying which features are important. With this interpretation, we do not expect $\theta$ to change across $\lambda$ values.

When $\theta$ is treated as a hyperparameter, we noted that the time needed for a grid search to find its optimal value grows exponentially with the number of sources of feature information. To avoid this computational burden, we suggested a descent algorithm for $\theta$ based on its gradient with respect to the fwelnet objective function (Step 3(a) and 3(b) in Algorithm \ref{alg_fwelnet}). There are other methods for hyperparameter optimization such as random search (e.g. \cite{Bergstra2012}) or Bayesian optimization (e.g. \cite{Snoek2012}) that could be applied to this problem.

One could consider $\theta$ as an argument of the fwelnet objective function to be minimized over jointly with $\beta$. One benefit of this approach is that it gives us a theoretical connection to the group lasso (Section \ref{sec:glasso}). However, we will obtain different estimates of $\theta$ for each value of the hyperparameter $\lambda$, which may be undesirable for interpretation. The objective function is also not jointly convex in $\theta$ and $\beta$, meaning that different minimization algorithms could end up at different local minima. In our attempts to make this approach work (see Appendix \ref{sec:theta_parameter}), it did not fare as well in prediction performance and was computationally expensive. It remains to be seen if there is a computationally efficient algorithm which treats $\theta$ as a parameter to be optimized for each $\lambda$ value. 

\end{itemize}

An R language package {\tt fwelnet} which implements our method is available at \url{https://www.github.com/kjytay/fwelnet}.

\medskip

{\bf Acknowledgements:} Nima Aghaeepour was supported by the Bill \& Melinda Gates Foundation (OPP1189911, OPP1113682), the National Institutes of Health (R01AG058417, R01HL13984401, R61NS114926, KL2TR003143), the Burroughs Wellcome Fund and the March of Dimes Prematurity Research Center at Stanford University. Trevor Hastie was partially supported by the National Science Foundation (DMS-1407548 and IIS1837931) and the National Institutes of Health (5R01 EB001988-21). Robert Tibshirani was supported by the National Institutes of Health (5R01 EB001988-16) and the National Science Foundation (19 DMS1208164).

\appendix
\section{Alternative algorithm with $\theta$ as a parameter}\label{sec:theta_parameter}

Assume that $\by$ and the columns of $\bX$ are centered so that $\hat{\beta}_0 = 0$ and we can ignore the intercept term in the rest of the discussion. If we consider $\theta$ as an argument of the objective function, then we wish to solve

\begin{align*}
(\hat{\beta}, \hat{\theta}) &= \underset{\beta, \theta}{\text{argmin}} \:\: J_{\lambda, \alpha}(\beta, \theta) \\
&= \underset{\beta, \theta}{\text{argmin}} \:\: \frac{1}{2}\|\by -\bX \beta\|_2^2+ \lambda \sum_{j=1}^p w_j(\theta) \left[\alpha |\beta_j | + \frac{1 - \alpha}{2} \beta_j^2 \right].
\end{align*}

$J$ is not jointly convex $\beta$ and $\theta$, so reaching a global minimum is a difficult task. Instead, we content ourselves with reaching a local minimum. A reasonable approach for doing so is to alternate between optimizing $\beta$ and $\theta$: the steps are outlined in Algorithm \ref{alg1}.

\begin{algorithm}
\caption{ \em Minimizing the fwelnet objective function via alternating minimization}
\label{alg1}
\begin{enumerate}
\item Select a value of $\alpha \in [0, 1]$ and a sequence of $\lambda$ values $\lambda_1 > \ldots > \lambda_m$.

\item For $i = 1, \ldots, m$:

		\begin{enumerate}
		\item Initialize $\beta^{(0)}(\lambda_i)$ at the elastic net solution for $\lambda_i$. Initialize $\theta^{(0)} = \bzero$.
		
		\item For $k = 0, 1, \ldots$ until convergence:
				\begin{enumerate}
				\item Fix $\beta = \beta^{(k)}$, update $\theta^{(k+1)}$ via gradient descent. That is, set $\Delta \theta = \left. \dfrac{\partial J_{\lambda_i, \alpha}}{\partial \theta} \right|_{\beta = \beta^{(k)}, \theta = \theta^{(k)}}$ and update $\theta^{(k+1)} = \theta^{(k)} - \eta \Delta \theta$, where $\eta$ is the step size computed via backtracking line search to ensure that $J_{\lambda_i, \alpha} \left(\beta^{(k)}, \theta^{(k+1)} \right) < J_{\lambda_i, \alpha} \left(\beta^{(k)}, \theta^{(k)} \right)$.
				
				\item Fix $\theta = \theta^{(k+1)}$, update $\beta^{(k+1)}$ by solving the elastic net with updated penalty factors $w_j (\theta^{(k+1)})$.
				\end{enumerate}
		
		\end{enumerate}
\end{enumerate}

\end{algorithm}

Unfortunately, Algorithm \ref{alg1} is slow due to repeated solving of the elastic net problem in Step 2(b)ii for each $\lambda_i$. The algorithm does not take advantage of the fact that once $\alpha$ and $\theta$ are fixed, the elastic net problem can be solved quickly for an entire path of $\lambda$ values. We have also found that Algorithm \ref{alg1} does not predict as well as Algorithm \ref{alg_fwelnet} in our simulations.

\appendix
\section{Details on simulation study in Section \ref{sec:sim}}\label{sec:simdetails}

\subsection{Setting 1: Noisy version of the true $\beta$}\label{sec:simdetails_1}

\begin{enumerate}
\item Set $n = 100$, $p = 50$, $\beta \in \bbR^{50}$ with $\beta_j = 2$ for $j = 1, \ldots, 5$, $\beta_j = -1$ for $j = 6, \ldots, 10$, and $\beta_j = 0$ otherwise.

\item Generate $x_{ij} \stackrel{i.i.d.}{\sim} \calN(0, 1)$ for $i = 1, \ldots, n$ and $j = 1, \ldots, p$.

\item For each $SNR_y \in \{ 0.5, 1, 2\}$ and $SNR_Z \in \{ 0.5, 2, 10 \}$:

		\begin{enumerate}
		\item Compute $\sigma_y^2 = \left(\sum_{j=1}^p \beta_j^2 \right) / SNR_y$.
		\item Generate $y_i = \sum_{j=1}^p x_{ij} \beta_j + \eps_i$, where $\eps_i \stackrel{i.i.d.}{\sim} \calN(0, \sigma_y^2)$ for $i = 1, \ldots, n$.
		\item Compute $\sigma_Z^2 = \text{Var} (|\beta|) / SNR_Z$.
		\item Generate $z_j = |\beta_j| + \eta_j$, where $\eta_j \stackrel{i.i.d.}{\sim} \calN(0, \sigma_Z^2)$. Append a column of ones to get $\bZ \in \bbR^{p \times 2}$.
		\end{enumerate}

\end{enumerate}

\subsection{Setting 2: Grouped data setting}

\begin{enumerate}
\item Set $n = 100$, $p = 150$.

\item For $j = 1, \ldots, p$ and $k = 1, \ldots 15$, set $z_{jk} = 1$ if $10(k-1) < j \leq 10k$, $z_{jk} = 0$ otherwise.

\item Generate $\beta \in \bbR^{150}$ with $\beta_j = 3$ or $\beta_j = -3$ with equal probability for $j = 1, \ldots, 10$, $\beta_j = 0$ otherwise.

\item Generate $x_{ij} \stackrel{i.i.d.}{\sim} \calN(0, 1)$ for $i = 1, \ldots, n$ and $j = 1, \ldots, p$.

\item For each $SNR_y \in \{ 0.5, 1, 2\}$:

		\begin{enumerate}
		\item Compute $\sigma_y^2 = \left(\sum_{j=1}^p \beta_j^2 \right) / SNR_y$.
		\item Generate $y_i = \sum_{j=1}^p x_{ij} \beta_j + \eps_i$, where $\eps_i \stackrel{i.i.d.}{\sim} \calN(0, \sigma_y^2)$ for $i = 1, \ldots, n$.
		\end{enumerate}

\end{enumerate}

\subsection{Setting 3: Noise variables}

\begin{enumerate}
\item Set $n = 100$, $p = 100$, $\beta \in \bbR^{100}$ with $\beta_j = 2$ for $j = 1, \ldots, 10$, and $\beta_j = 0$ otherwise.

\item Generate $x_{ij} \stackrel{i.i.d.}{\sim} \calN(0, 1)$ for $i = 1, \ldots, n$ and $j = 1, \ldots, p$.

\item For each $SNR_y \in \{ 0.5, 1, 2\}$:

		\begin{enumerate}
		\item Compute $\sigma_y^2 = \left(\sum_{j=1}^p \beta_j^2 \right) / SNR_y$.
		\item Generate $y_i = \sum_{j=1}^p x_{ij} \beta_j + \eps_i$, where $\eps_i \stackrel{i.i.d.}{\sim} \calN(0, \sigma_y^2)$ for $i = 1, \ldots, n$.
		\item Generate $z_{jk} \stackrel{i.i.d.}{\sim} \calN(0, 1)$ for $j = 1, \ldots, p$ and $k = 1, \ldots 10$. Append a column of ones to get $\bZ \in \bbR^{p \times 11}$.
		\end{enumerate}

\end{enumerate}

\section{Proof of Theorem \ref{thm:glasso}}\label{sec:glasso_proof}

First note that if feature $j$ belongs to group $k$, then $\bz_j^T \theta = \theta_k$, and its penalty factor is
\begin{equation*}
w_j(\theta) = \dfrac{\sum_{\ell = 1}^p f(\bz_\ell^T \theta)}{pf(\bz_j^T \theta)} = \dfrac{\sum_{\ell = 1}^p f(\theta_\ell)}{pf(\theta_k)} = \dfrac{\sum_{\ell=1}^K p_\ell f(\theta_\ell)}{pf(\theta_k)},
\end{equation*}

where $p_\ell$ denotes the number of features in group $\ell$. Letting $v_k = \dfrac{f(\theta_k)}{\sum_{\ell=1}^K p_\ell f(\theta_\ell)}$ for $k = 1, \ldots, K$, minimizing the fwelnet objective function \eqref{eqn:fwelnet} over $\beta$ and $\theta$ reduces to

\begin{align*}
\underset{\beta, \theta}{\minimize} \quad &\frac{1}{2} \ltwo{\by - \bX \beta}^2 + \frac{\lambda}{p} \sum_{k = 1}^K \frac{1}{v_k} \left[ \alpha \lone{\beta^{(k)}} + \frac{1-\alpha}{2} \ltwo{\beta^{(k)}}^2 \right].
\end{align*}

For fixed $\beta$, we can explicitly determine the $v_k$ values which minimize the expression above. By the Cauchy-Schwarz inequality,

\begin{align}
\frac{\lambda}{p} \sum_{k = 1}^K \frac{1}{v_k} \left[ \alpha \lone{\beta^{(k)}} + \frac{1-\alpha}{2} \ltwo{\beta^{(k)}}^2 \right] &= \frac{\lambda}{p} \left( \sum_{k = 1}^K \frac{1}{v_k} \left[ \alpha \lone{\beta^{(k)}} + \frac{1-\alpha}{2} \ltwo{\beta^{(k)}}^2 \right] \right) \left( \sum_{k=1}^K p_k v_k \right) \nonumber \\ 
&\geq \frac{\lambda}{p} \left( \sum_{k = 1}^K \sqrt{p_k \left[ \alpha \lone{\beta^{(k)}} + \frac{1-\alpha}{2} \ltwo{\beta^{(k)}}^2 \right]} \right)^2. \label{eqn:glasso-proof1}
\end{align}

Note that equality is attainable for \eqref{eqn:glasso-proof1}: letting $a_k = \sqrt{\frac{ \left[ \alpha \lone{\beta^{(k)}} + \frac{1-\alpha}{2} \ltwo{\beta^{(k)}}^2 \right] }{p_k}}$, equality occurs when there is some $c \in \bbR$ such that
\begin{align*}
c \cdot \frac{1}{v_k} \left[ \alpha \lone{\beta^{(k)}} + \frac{1-\alpha}{2} \ltwo{\beta^{(k)}}^2 \right] &= p_k v_k &\text{for all } k, \\
v_k &=  \sqrt{c} a_k &\text{for all } k.
\end{align*}

Since $\sum_{k=1}^K p_k v_k = 1$, we have $\sqrt{c} = \dfrac{1}{\sum_{k=1}^K p_k a_k}$, giving $v_k = \dfrac{a_k}{\sum_{k=1}^K p_k a_k}$ for all $k$. A solution for this is $f(\theta_k) = a_k$ for all $k$, which is feasible for $f$ having range $[0, \infty)$. (Note that if $f$ only has range $(0, \infty)$, the connection still holds if $\lim_{x \rightarrow -\infty} f(x) = 0$ or $\lim_{x \rightarrow +\infty} f(x) = 0$: the solution will just have $\theta = +\infty$ or $\theta = -\infty$.)

Thus, the fwelnet solution is
\begin{align}\label{eqn:glasso-proof-langrange}
&\underset{\beta}{\text{argmin}} \quad \frac{1}{2} \ltwo{\by - \bX \beta}^2 + \frac{\lambda}{p} \left( \sum_{k = 1}^K \sqrt{p_k \left[ \alpha \lone{\beta^{(k)}} + \frac{1-\alpha}{2} \ltwo{\beta^{(k)}}^2 \right]} \right)^2.
\end{align}

Writing in constrained form, \eqref{eqn:glasso-proof-langrange} becomes minimizing $\frac{1}{2} \ltwo{\by - \bX \beta}^2$ subject to
\begin{align*}
\left( \sum_{k = 1}^K \sqrt{p_k \left[ \alpha \lone{\beta^{(k)}} + \frac{1-\alpha}{2} \ltwo{\beta^{(k)}}^2 \right]} \right)^2 &\leq C \text{ for some constant } C, \\
\sum_{k = 1}^K \sqrt{p_k \left[ \alpha \lone{\beta^{(k)}} + \frac{1-\alpha}{2} \ltwo{\beta^{(k)}}^2 \right]} &\leq \sqrt{C}.
\end{align*}

Converting back to Lagrange form again, there is some $\lambda' \geq 0$ such that the fwelnet solution is
\begin{align*}
&\underset{\beta}{\text{argmin}} \quad \frac{1}{2} \ltwo{\by - \bX \beta}^2 + \lambda' \sum_{k = 1}^K \sqrt{p_k \left[ \alpha \lone{\beta^{(k)}} + \frac{1-\alpha}{2} \ltwo{\beta^{(k)}}^2 \right]}.
\end{align*}

Setting $\alpha = 0$ and $\alpha = 1$ in the expression above gives the desired result.

\section{Details on simulation study in Section \ref{sec:multitask}}\label{sec:sim_multitask}

\begin{enumerate}
\item Set $n = 150$, $p = 50$.

\item Generate $\beta_1 \in \bbR^{50}$ with 
\begin{equation*}
\beta_{1,j} = \begin{cases} 5 \text{ or } -5 \text{ with equal probability} &\text{for } j = 1, \dots, 5, \\
2 \text{ or } -2 \text{ with equal probability} &\text{for } j = 6, \dots, 10, \\
0 &\text{otherwise.} \end{cases}
\end{equation*}

\item Generate $\beta_2 \in \bbR^{50}$ with 
\begin{equation*}
\beta_{2,j} = \begin{cases} 5 \text{ or } -5 \text{ with equal probability} &\text{for } j = 1, \dots, 5, \\
2 \text{ or } -2 \text{ with equal probability} &\text{for } j = 11, \dots, 15, \\
0 &\text{otherwise.} \end{cases}
\end{equation*}

\item Generate $x_{ij} \stackrel{i.i.d.}{\sim} \calN(0, 1)$ for $i = 1, \ldots, n$ and $j = 1, \ldots, p$.

\item Generate response 1, $\by_1 \in \bbR^{150}$, in the following way:

		\begin{enumerate}
		\item Compute $\sigma_1^2 = \left(\sum_{j=1}^p \beta_{1,j}^2 \right) / 0.5$.
		\item Generate $y_{1,i} = \sum_{j=1}^p x_{ij} \beta_{1,j} + \eps_{1,i}$, where $\eps_{1,i} \stackrel{i.i.d.}{\sim} \calN \left(0, \sigma_1^2 \right)$ for $i = 1, \ldots, n$.
		\end{enumerate}

\item Generate response 2, $\by_2 \in \bbR^{150}$, in the following way:

		\begin{enumerate}
		\item Compute $\sigma_2^2 = \left(\sum_{j=1}^p \beta_{2,j}^2 \right) / 1.5$.
		\item Generate $y_{2,i} = \sum_{j=1}^p x_{ij} \beta_{2,j} + \eps_{2,i}$, where $\eps_{2,i} \stackrel{i.i.d.}{\sim} \calN \left(0, \sigma_2^2 \right)$ for $i = 1, \ldots, n$.
		\end{enumerate}

\end{enumerate}

\bibliographystyle{agsm}
\bibliography{fwelnet}

\end{document}